# Non-linear equations for electron waves in Maxwellian low-collision ion-electron plasmas[1]

V. N. Soshnikov [2]

Plasma Physics Dept.,
All-Russian Institute of Scientific and Technical Information
of the Russian Academy of Sciences
*(VINITI, Usievitcha 20, 125315 Moscow, Russia)*

*The before described general principles and methodology of calculating electron wave propagation in homogeneous isotropic half-infinity slab of Maxwellian plasma with indefinite but in principal value sense taken integrals in characteristic equations, and the use of 2D Laplace transform method are applied to an evaluation of collision damping decrements of plane electron longitudinal and transverse waves. Damping decrement tends to infinity when the wave frequency tends to electron Langmuir frequency $\omega \to \omega_L + 0$. We considered recurrent relations for amplitudes of the overtones which form in their sum the all solution of the plasma wave non-linear equations including collision damping and quadratic (non-linear) terms. Collisionless damping at $\omega > \omega_L$ is possible only in non-Maxwellian plasmas.*

**Introduction**

In the presented work we intend mainly demonstrate the of principle approach to calculation wave processes in plasma, based in principle on refusal from the so-called Landau damping theory. Therefore having in mind some general issues we have considered the simplest cases of the well known boundary problem with periodic boundary electrical field at coordinate $x = 0$ of the half-infinite homogeneous isotropic fully ionized plasma slab, with zero initial conditions, considering propagation of both electron longitudinal (electrostatic) and transverse (electromagnetic and coupled with them electron) waves.

As it is well known, the solutions contain indefinite logarithmically divergent integrals. Appearance of these integrals ought to be considered as an evidence of information lack in original (kinetic and Maxwell) equations, so one must add some additional physical conditions of the problem which also would define the only way of taking such integrals. In the case of half-infinite plasma slab such evident conditions are the absence of the fastest (among existing) backward waves and also so-called kinematic waves, not connected with the boundary electric field. The ubiquitous way of taking Landau contour integrals with analytical continuation of integrand function of the real value velocity $v_x$ in its poles on the real axis into complex plane results in the mathematical solutions which don't satisfy the above mentioned additional physical conditions and cannot be therefore acknowledged as correct ones [1], [2], [3].

Since one has in mind here that the background Maxwellian function $f_0(v)$ keeps its form at the boundary of plasma slab $x = 0$, it implies either the presence of an infinite $(\pm\infty)$ extensive plasma with the boundary field $E_0 \exp(i\omega t)$ applied in some plane $x = 0$ of the plasma volume or the presence of an ideally reflecting boundary plane.

---

[1] It is English version of the paper published in the Russian journal "Engineering Physics", 2005, n. 2, p. 42.
[2] Krasnodarskaya str., 51-2-168, Moscow 109559, Russia. E-mail: vikt3363@yandex.ru .

The general form of the problem equations is drastically simplified by using some clear approximations in the form of introducing some effective constant value of integrand variable $v_{xeff}$ what allows to constitute at least qualitative idea about really existing main dependences [4]. The resulting solutions of dispersion equations and damping decrements can be then corrected by introducing some correcting coefficients of the order $\gamma \sim 1,5 \div 2$ (see [10]), what however doesn't lead to qualitative change of the main consequences. Preliminary results are presented in electron versions of papers [2], [5], [6], [7], [8], [10], [12].

In this work results on collision damping are unified and generalized with the only view point including some precision issues and discussion on some iteration procedures of calculation of collision damping and approximations in overtone expansion with cutting off non-physical inadmissible values of distribution function.

The following description is based on the preliminary solving in the above manner the well known traditional original equations of plasma waves

$$\frac{\partial f_1}{\partial t} + v_x \frac{\partial f_1}{\partial x} - \frac{|e|E_x(x,t)}{m}\frac{\partial f_0}{\partial v_x} = \frac{|e|E_x(x,t)}{m}\frac{\partial f_1}{\partial v_x} + Q, \tag{1}$$

$$\frac{\partial E_x(x,t)}{\partial x} = -4\pi|e|n_e \int f_1 d\vec{v} \tag{2}$$

(self-consistent Vlasov equations for longitudinal waves) and

$$\frac{\partial f_1}{\partial t} + v_x \frac{\partial f_1}{\partial x} - \frac{|e|E_z(x,t)}{m}\frac{\partial f_0}{\partial v_z} = \frac{|e|E_z(x,t)}{m}\frac{\partial f_1}{\partial v_z} + Q, \tag{3}$$

$$\frac{\partial^2 E_z(x,t)}{\partial x^2} - \frac{1}{c^2}\frac{\partial^2 E_z(x,t)}{\partial t^2} + \frac{4\pi|e|n_e}{c^2}\frac{\partial}{\partial t}\int v_z f_1 d\vec{v} = 0 \tag{4}$$

(equations of propagation of transverse electromagnetic waves) where

$$f_0 = \left(\frac{m}{2\pi k_B T}\right)^{3/2} e^{-\frac{mv^2}{2k_B T}}. \tag{5}$$

The right hand sides in expressions (1) and (3) contain usually assumed to be small quadratic non-linear terms and the Coulomb collisions integral of electron impacts with immovable ions of the standard form (see [9])

$$Q(\vec{v},x,t) \equiv \hat{Q}(f_1) \simeq \frac{2\pi e^4 L Z_i^2 n_i}{m^2}\frac{\partial}{\partial v_i}\left(\frac{v^2 \delta_{ij} - v_i v_j}{v^3} \cdot \frac{\partial f_1(\vec{v},x,t)}{\partial v_j}\right), \tag{6}$$

where $\hat{Q}(f_1)$ is linear differential operator applied to the function $f_1$; $n_e$ and $n_i$ correspondingly are electron and ion densities; $k_B$ is Boltzmann constant; $m$ is electron mass; $L$ is so-called Coulomb logarithm; further symbols $|e|$, $E_x$, $E_z$ are replaced with $e$ and $E$; the waves propagate along coordinate $x$; distribution function is represented as a sum of normalized to unity Maxwellian function $f_0$ and of small perturbation $f_1$, $|f_1| \ll f_0$. Linearization of these equations reduces to neglecting the small r.h.s. quadratic terms of equations (1) and (3).

For to obtain asymptotical solutions one uses classical Laplace transform method with integral

representations

$$E(x,t) = \frac{1}{(2\pi i)^2} \int_{\sigma_1^-}^{\sigma_1^+} \int_{\sigma_2^-}^{\sigma_2^+} E_{p_1 p_2} e^{p_1 t + p_2 x} dp_1 dp_2, \qquad (7)$$

$$f_1(\vec{v},x,t) = \frac{1}{(2\pi i)^2} \int_{\sigma_1^-}^{\sigma_1^+} \int_{\sigma_2^-}^{\sigma_2^+} f_{p_1 p_2} e^{p_1 t + p_2 x} dp_1 dp_2, \qquad (8)$$

$$\frac{\partial E(x,t)}{\partial x} = \frac{1}{(2\pi i)^2} \int_{\sigma_1^-}^{\sigma_1^+} \int_{\sigma_2^-}^{\sigma_2^+} \left( p_2 E_{p_1 p_2} - E_{p_1} \right) e^{p_1 t + p_2 x} dp_1 dp_2, \qquad (9)$$

$$\frac{\partial f_1(\vec{v},x,t)}{\partial x} = \frac{1}{(2\pi i)^2} \int_{\sigma_1^-}^{\sigma_1^+} \int_{\sigma_2^-}^{\sigma_2^+} \left( p_2 f_{p_1 p_2} - f_{p_1} \right) e^{p_1 t + p_2 x} dp_1 p_2, \qquad (10)$$

$$\frac{\partial f_1(\vec{v},x,t)}{\partial t} = \frac{1}{(2\pi i)^2} \int_{\sigma_1^-}^{\sigma_1^+} \int_{\sigma_2^-}^{\sigma_2^+} \left( p_1 f_{p_1 p_2} - f_{p_2} \right) e^{p_1 t + p_2 x} dp_1 dp_2, \qquad (11)$$

$$\frac{\partial^2 E(x,t)}{\partial x^2} = \frac{1}{(2\pi i)^2} \int_{\sigma_1^-}^{\sigma_1^+} \int_{\sigma_2^-}^{\sigma_2^+} (p_2^2 E_{p_1 p_2} - p_2 E_{p_1} - F_{p_1}) e^{p_1 t + p_2 x} dp_1 dp_2, \qquad (12)$$

$$\frac{\partial^2 E(x,t)}{\partial t^2} = \frac{1}{(2\pi i)^2} \int_{\sigma_1^-}^{\sigma_1^+} \int_{\sigma_2^-}^{\sigma_2^+} \left( p_1^2 E_{p_1 p_2} - p_1 E_{p_2} - F_{p_2} \right) e^{p_1 t + p_2 x} dp_1 dp_2, \qquad (13)$$

where $\sigma_{1,2}^\pm = \sigma_{1,2} \pm i \cdot \infty$, $f_{p_1}$, $f_{p_2}$, $F_{p_1}$, $E_{p_1}$ and $E_{p_2}$, correspondingly, are Laplacian images of the functions

$$f_1(\vec{v},0,t), \ f_1(\vec{v},x,0), \ \left.\frac{\partial E(x,t)}{\partial x}\right|_{x=0}, \ \left.\frac{\partial E(x,t)}{\partial t}\right|_{t=0}, \ E(0,t), \ E(x,0). \qquad (14)$$

In a considered further self-consistent boundary problem with the given periodic boundary field other boundary conditions are not arbitrary, and it ought to point out a way of their finding.

Taking into consideration for the most part demonstrative aim of this work we use further as it has been made already in the works [4], [5], [6], [7] approximation of replacing $v_x \to v_{xeff} = const$ [4] with $v_{xeff} \approx (1.5 \div 2)\sqrt{\overline{v_x^2}}$, what drastically simplifies calculation of integrals of the type

$$\int_{-\infty}^{+\infty} \frac{e^{-\frac{mv^2}{2k_B T}}}{p_1 + v_x p_2} dv_x \equiv \int_0^{+\infty} \frac{e^{-\frac{mv^2}{2k_B T}} 2 p_1 dv_x}{p_1^2 - v_x^2 p_2^2} \approx \sqrt{\frac{2\pi k_B T}{m}} \frac{p_1}{p_1^2 - v_x^2 p_2^2}, \qquad (15)$$

where

$$\overline{v_x^2} = \frac{k_B T}{m} \qquad (16)$$

ought to be replaced further with the suitable more precise value $v_{xeff}^2$.

Thus it corresponds to an approximate calculation of the principal sense value of indefinite integrals.

For low collision plasmas with small $|Q|$ one could use further an iteration method beginning on calculation of the first iteration.

The boundary field can be given either as

$$E(0,t) = E_1 \exp(i\omega t) \tag{17}$$

or

$$E(0,t) = E_1 \cos \omega t \tag{18}$$

(see [8]), what has no importance at finding Laplacian dispersion equation poles.

In the work [8] there is presented solution of the non-linear collisionless equation of plane waves propagation in the form of an expansion in real value overtones with decreasing amplitudes. Analogues expansions one can obtain at the presence of collision damping using series of complex value and its complex conjugated overtones of the form $\sim \exp[in(\omega t - k_0 x + \varphi) - \delta x]$ of the order $n = 1, 2, \ldots$, where $k_0$ is real value part of the wave number, and $\delta$ is damping decrement. Owing to the presence of quadratic term, this expansion will contain also overtones of the order $n$ in the form $\sim \exp[i(q+r)(\omega t - k_0 x + \varphi) - \delta x]$ with the whole-number positive and negative $q$, $r \neq 0$, $q + r \neq 0$, $q + r = n$ with amplitudes which are proportional $(E_1)^{|q|+|r|}$, thus these terms give small corrections for the highest powers of $E_1$ (at $|q|+|r| > n$) to other overtones of the order $n$. So, in the dispersion equation corresponding to $n = 1$ all $E_1$ are cancelled, the corrections begin only with the term $\sim E_1^2$, $n \geq 2$. Correspondingly, for terms with amplitudes $\sim E_1^2$, $E_1^3$, ..., corrections are subsequently of the order $E_1^4$, $E_1^5$, .... Thus, the general having physical sense solution can be represented in the form of a sum of products of pairs complex conjugated series of terms correspondingly with $n > 0$ at a presence of mixed terms with amplitudes of diverse orders $n$ (see further Section 4).

**Longitudinal waves**

After proceeding Laplace transform Eqs. (1), (2) take the form

$$f_{p_1 p_2} = \frac{\frac{e}{m} \frac{\partial f_0}{\partial v_x} E_{p_1 p_2} + f_{p_2} + v_x f_{p_1}}{p_1 + v_x p_2}, \tag{19}$$

$$E_{p_1 p_2} = \frac{1}{p_2 G(p_1 p_2)} \left( E_{p_1} + \int \frac{f_{p_2} + v_x f_{p_1}}{p_1 + v_x p_2} d\vec{v} \right), \tag{20}$$

$$G(p_1 p_2) = 1 + \frac{\omega_L^2}{p_2} \int \frac{\partial f_0 / \partial v_x}{p_1 + v_x p_2} d\vec{v} \approx 1 + \frac{\omega_L^2}{p_1^2 - v_x^2 p_2^2}. \tag{21}$$

Asymptotical solutions are defined by simultaneous pair (in $p_1$, $p_2$) poles of Laplace images

$f_{p_1 p_2}$, $E_{p_1 p_2}$ with amplitudes which are defined by the sum of residua of all pair poles. For the fixed given boundary frequency and $E(0,t) = E_1 \exp i\omega t$ one has

$$E_{p_1} = \frac{E_1}{p_1 - i\omega} \tag{22}$$

with a pole

$$p_1^{(0)} = i\omega. \tag{23}$$

The second pole $p_2^0$ is defined by equation

$$G\left(i\omega, p_2^{(0)}\right) = 0, \tag{24}$$

that is, since $p_2 \equiv ik$, for the value $k = k(\omega)$ in the pole $p_2^{(0)}$ one has

$$k = \pm \omega \frac{\sqrt{1 - \omega_L^2/\omega^2}}{\sqrt{v_x^2}}. \tag{25}$$

At $E(0,t) = E_1 \exp i\omega t$ it is natural to expect that $f_1(0,t)$ can be proportional to $\exp i\omega t$ with the pole $p_1^{(0)} = i\omega$. The term $\sim f_{p_2}$ corresponding to zero initial conditions doesn't contain above-mentioned poles in $p_1$ and can be omitted in this setting of the boundary problem. Thus, asymptotical solution for the field $E(x,t)$ contains a sum of at least two types of terms with exponent factors $\exp(i\omega t \pm ikx)$, that is forward and non-physical backward waves in the given half-slab boundary problem. In the case of $E(0,t) = E_1 \cos \omega t$ the number of waves twists to the types $\pm(i\omega t - ikx)$ and $\pm(i\omega t + ikx)$.

Substituting expression for $E_{p_1 p_2}$ from Eq. (20) to Eq. (19) one obtains

$$f_{p_1 p_2} = \frac{1}{p_1 + v_x p_2} \left[ v_x f_{p_1} + \frac{e}{m} \frac{\partial f_0}{\partial v_x} \frac{1}{p_2 G(p_1 p_2)} \left( E_{p_1} + \int \frac{v_x f_{p_1}}{p_1 + v_x p_2} d\vec{v} \right) \right], \tag{26}$$

from which it follows a possibility of existing so-called kinematic ("ballistic") waves which correspond to simultaneous (pair) poles

$$p_1 = i\omega, \quad p_2 = -i\omega/v_x. \tag{27}$$

Theoretically existing solution in the form of kinematic wave $\sim \exp\left(i\omega t - i\frac{\omega}{v_x} x\right)$ is however non-physical since this wave propagates with velocity $v_x$, and relative to the sign of $v_x$ in both directions, as forward, as well as backward, but in the one-sided boundary problem there are no physical sources which would produce distorting symmetric relative $\pm v_x$ background Maxwellian distribution function with arising wave movement in backward direction. And what is more, even in the general case of plasma slab with two boundaries it ought to exclude appearing mathematically acceptable for Eqs. (1) and (3) but unphysical in the case of charged particles (inacceptable for Eqs. (2) and (4)) for both forward and backward kinematic waves and oscillations as not connected with the given boundary electrical field and being in contradiction with the taken by definition background Maxwellian distribution function at the absence of electrical field.

Besides that, kinematic waves, since they are not supported by electrical field, must some time or other fade out owing to existing even in the so-called "collisionless" plasma as much as one wants rare collisions. But all this doesn't exclude certainly a possibility of supporting some kinematic waves on account of outward non-electrical forces but also with addition to the field $E((0,t)$ (and probably to $E(x,t)$) of plasma charged particles some outward electric field.

Kinematic constituent of the function $f_1(x,t,v,v_x)$ has the form

$$f_1^{kinem} = e^{i\omega t}e^{-i\omega x/v_x}F(v,v_x,E_1,\omega), \tag{28}$$

and its substitution into the kinetic equation leads to the sought for condition of absence of kinematic waves $F(v,v_x,E_1,\omega)=0$ or, in an expansion form, to condition of zero residuum

$$\left\{(p_1-i\omega)\left[v_x f_{p_1} + \frac{e}{m}\frac{\partial f_0/\partial v_x}{p_2 G(p_1,p_2)}\left(E_{p_1} + \int\frac{v_x f_{p_1}}{p_1+v_x p_2}\right)\right]\right\}\Bigg|_{\substack{p_1\to i\omega\\p_2\to -i\omega/v_x}} = 0. \tag{29}$$

This integral equation determines in the main dependence $f_{p_1}$ on $\vec{v}$, correspondingly the boundary distribution function $f_1(0,t,v,v_x)$ with its proportionality $f_1 \sim E_1$.[3]

From equations (26) and (20) it follows directly that the condition of absence of non-physical backward waves both for integrand inclusion $f_{p_1}(v,v_x)$ and $E_{p_1}$ (that is $f_1(0,t,v,v_x)$ and $E(0,t)$) is implemented at the same proportionality fraction between $E_1$ and $f_{p_1}$ in integrands of (20) and (26), but in the general case this relation will be other than in linear integral equation (29).

One can satisfy simultaneously the conditions of absence as kinematic, as well as backward waves only at the natural assumption of a bound of the constituent of electric field $E_1$ at the boundary with corresponding them the bound of a surface charge density with different from (29) relation of $E_1$ and $f_1$. In this manner the changed relation ought to be determined from polarization considerations of connection of the surface charge density with applied outward field. In other words it is assumed that the total boundary field is a sum of the two harmonic components: the given outward field mainly supporting wave process, and some intrinsic opposite polarization field preventing formation of kinematic waves.

Special consideration ought to be proceeded in cases which we here don't analyze: when the background distribution function is only approximately half-Maxwellian (either at partly absorbing or partly reflecting boundary).

In the case of non-kinematical wave the term $v_x f_{p_1}$ in Eq. (26) fall out at transition from Laplacian image to original, and the function $f_{p_1}$ remains present only in the integrand by velocity $v_x$.

---

[3] Vice versa, the picking out some function $f_{p_1}(v,v_x)$, correspondingly $f_1(0,t,v,v_x)$, should define from Eq. (29) a boundary value $E_{p_1}$ and $E(0,t)$, however really the form of $f_{p_1}(v,v_x)$ is strictly constrained by the requirement of independence $E_{p_1}$ of $v,v_x$.

Therefore, following (26), a linear approach with the boundary condition (17) takes a simple asymptotic form

$$f_1(\vec{v}, x, t) \sim \frac{\partial f_0/\partial v_x}{\omega - k v_x} e^{i\omega t \pm i k x}. \tag{30}$$

Thus, further one can be distracted from Eq. (29), which defines the form of $f_{p_1}(\vec{v}, v_x)$ in the general case.

Divergence of the function $f_1$ at $\omega = k v_x$ is caused by inapplicability of kinetic equation in this region, and can be removed out according the proposed in [8] cutting off $f_1$ owing to a requirement of positivity of the real-valued total distribution function $f_0 + \frac{1}{2}(f_1 + f_1^*)$, where $f_1^*$ is complex conjugated to $f_1$.

Substitution of collisionless solution (30) in the standard integral of Coulomb collisions (6), after neglecting terms $O(m_e/m_i)$, $O(\overline{v}_i/\overline{v}_e)$, $O(m_e T_e/m_i T_i)$ and transition to the Laplace image $Q_{p_1 p_2}$ leads to an expression [5]

$$E_{p_1 p_2} = \frac{E_{p_1}}{p_2} + \omega_L^2 \frac{m}{k_B T} \left(\frac{m}{2\pi k_B T}\right)^{3/2} \frac{E_{p_1 p_2}}{p_2} \int \frac{v_x e^{-\frac{mv^2}{2k_B T}}}{p_1 + v_x p_2} d\vec{v} + 8\pi^2 e^6 L \frac{Z^2 n_i n_e}{m^2} \left(\frac{m}{2\pi k_B T}\right)^{3/2} \int d\vec{V} e^{-\frac{m_i V^2}{2k_B T_i}} \times$$

$$\times \int d\vec{v} \frac{1}{(p_1 + v_x p_2)^2} \frac{u^2 \delta_{xl} - u_x u_l}{u^3} \frac{\partial}{\partial v_l} \left[\frac{v_x \exp\left(-\frac{mv^2}{2k_B T}\right)}{k_B T(p_1 + p_2 v_x)}\right], \tag{31}$$

where $\vec{u} \equiv \vec{v} - \vec{V}$, $\vec{V}$ is ion velocity, and $u^2 \approx v^2$; $u_l \approx v_l$.

Going from integration limits $(-\infty, +\infty)$ in $v_x$ to limits $(0, +\infty)$, proceeding differentiations $\partial/\partial v_l$ in the integrand in the last term of r.h.s. of expression (31), and going from $v$, $v^2$, $v_x^2$ at integration to their mean values $v \to \sqrt{3 \overline{v_x^2}}$, $v^2 \to 3 \overline{v_x^2}$, $v_x^2 \to \overline{v_x^2}$, using procedures analogous to them in (15) one obtains

$$E_{p_1 p_2} = \frac{E_1}{p_2(p_1 - i\omega)} \frac{\left(p_1^2 - p_2^2 \overline{v_x^2}\right)^4}{\left(p_1^2 - p_2^2 \overline{v_x^2}\right)^4 + \omega_L^2 \left(p_1^2 - p_2^2 \overline{v_x^2}\right)^3 - \frac{16\pi^2 e^6 Z^2 L p_1 n_e n_i}{3\sqrt{3\overline{v_x^2}} k_B T m^2} \left[p_1^4 + \left(p_2^2 \overline{v_x^2}\right)^2 + 6 p_1^2 p_2^2 \overline{v_x^2}\right]}, \tag{32}$$

where $E_1 = const$ corresponds to an amplitude of a boundary field (with account for assumed boundary polarization effect).

At small $|\delta| \ll \sqrt{(\omega^2 - \omega_L^2)/\overline{v_x^2}}$ this function has poles

$$p_1 = i\omega; \quad p_2 = \pm i\sqrt{\frac{\omega^2 - \omega_L^2}{\overline{v_x^2}}} + \delta, \tag{33}$$

what after its substitution into (32) with account for the smallness of $|\delta|$ leads to a coordinate increment/decrement value

$$\delta \simeq \pm \frac{8\pi^2 e^6 Z^2 L n_e n_i}{3\sqrt{3} m (k_B T)^2} \left[ \frac{8\omega^2 (\omega^2 - \omega_L^2) + \omega_L^4}{\omega_L^6 \sqrt{1 - \omega_L^2/\omega^2}} \right], \tag{34}$$

where $\delta < 0$ corresponds to forward waves with $k < 0$ [5].
Some intriguing peculiarity of this solution appears as tending $\delta \to -\infty$ at $\omega \to \omega_L + 0$.

It was noted above that at a more precise calculation mean velocities values at replacing in integrands $v \to \overline{v}$, $v_x \to \overline{v}_x$, $v_x^2 \to \overline{v_x^2}$ etc. must further be replaced with corresponding constant values $v_{eff}$, $v_{xeff}$, $v_{xeff}^2$, …, as it has been considered in [4], [10]. These effective values exceed mean values. For example, in the case of collisionless plasma $v_{xeff}^2 \sim (2 \div 4)\overline{v_x^2}$ (more details about corrections in the case of a low-collision plasma are in the work [10]), however at replacing mean values with some effective ones in resulting formulas the general form of these ones doesn't change, since dependence of proportionality coefficients $\gamma_i$ between the mean and the effective velocity values on parametric wave number $k$ is very weak.

**Transverse waves**

Substituting into equations (3), (4) Laplacian representations (7) … (13) and neglecting at $p_1 = i\omega$ initial and boundary functions (14) which don't affect the form of dispersion equation for determining $p_2$, one obtains this equation in the form

$$p_2^2 - \frac{p_1^2}{c^2} + \frac{\omega_L^2 p_1}{c^2} \int v_z \frac{\partial f_0}{\partial v_z} \frac{d\vec{v}}{p_1 + v_x p_2} = 0. \tag{35}$$

Using approaches of the type (15) one obtains equation (35) in the form

$$p_2^2 - \frac{p_1^2}{c^2} \left( 1 + \frac{\omega_L^2}{p_1^2 - \overline{v_x^2}} \right) = 0 \tag{36}$$

with two solution branches at $\overline{v_x^2} \ll c^2$:

$$\left[ p_2^{(1)} \right]^2 \equiv -k^2 = -\frac{\omega^2}{2\overline{v_x^2}} \left[ 1 + \frac{\overline{v_x^2}}{c^2} - \sqrt{\left(1 + \frac{\overline{v_x^2}}{c^2}\right)^2 - 4\frac{\overline{v_x^2}}{c^2}\left(1 - \frac{\omega_L^2}{\omega^2}\right)} \right] \approx -\frac{\omega^2}{c^2}\left(1 - \frac{\omega_L^2}{\omega^2}\right); \tag{37}$$

$$\left[ p_2^{(2)} \right]^2 \equiv -k^2 = -\frac{\omega^2}{2\overline{v_x^2}} \left[ 1 + \frac{\overline{v_x^2}}{c^2} + \sqrt{\left(1 + \frac{\overline{v_x^2}}{c^2}\right)^2 - 4\frac{\overline{v_x^2}}{c^2}\left(1 - \frac{\omega_L^2}{\omega^2}\right)} \right] \approx -\frac{\omega^2}{\overline{v_x^2}}\left(1 + \frac{\overline{v_x^2} \omega_L^2}{2c^2 \omega^2}\right) \approx -\frac{\omega^2}{\overline{v_x^2}}. \tag{38}$$

Inclusion of the Coulomb collision integral $Q$ in the form (6) into (3) leads to a dispersion equation at $p_1 = i\omega$ in the first iteration approach for the distribution function

$$\left(p_1^2 - p_2^2 \overline{v_x^2}\right)^4 \left(p_2^2 - \frac{p_1^2}{c^2}\right) - \left(p_1^2 - p_2^2 \overline{v_x^2}\right)^3 \frac{p_1^2 \omega_L^2}{c^2} + \frac{4\pi e^4 \omega_L^2 Z^2 n_i L}{3\sqrt{3\overline{v_x^2}} mc^2 k_B T} p_1 \left[ p_1^6 - 3p_1^4 p_2^2 \overline{v_x^2} + \right.$$

$$\left. + 7p_1^2 \left(p_2^2 \overline{v_x^2}\right)^2 + 3\left(p_2^2 \overline{v_x^2}\right)^3 \right] = 0 . \tag{39}$$

At $p_2 = p_2^{(1)} + \delta$, $|\delta| \ll |p_2^{(1)}|$ one obtains for the first branch with forward wave $k < 0$ coordinate damping decrement

$$\delta \equiv \delta_1 \approx -\frac{2\pi e^4 L Z^2 n_i \omega_L^2}{3\sqrt{3\overline{v_x^2}} m k_B T c \omega^2 \sqrt{1 - \omega_L^2/\omega^2}} \tag{40}$$

with $\delta \to -\infty$ at $\omega \to \omega_L + 0$.

For the second branch, taking case $k < 0$ and proposing

$$\frac{\sqrt{\overline{v_x^2}} \omega_L^2}{2\omega c^2} \ll |\delta| \ll \frac{m}{\sqrt{\overline{v_x^2}}} \tag{41}$$

one obtains

$$\delta \equiv \delta_2 \approx -\left[ \frac{\pi e^4 L Z^2 n_i \omega^2}{3\sqrt{3} \left(\overline{v_x^2}\right)^2 m k_B T} \right]^{1/3} . \tag{42}$$

As in the case of longitudinal waves one can always obtain by a selection of self-consistent initial and boundary conditions (14) the solution which doesn't contain non-physical backward and kinematic waves, but now without any polarization or some other bound at the boundary.

About correction with replacing mean velocities in integrands with some effective values, see also [10].

**Non-linear equations of electron wave damping in low-collision plasmas**

The solution of electron plasma waves non-linear equations can be represented in the form of a sum of n-order overtones with decreasing amplitudes $\sim E_1^n$ (see [8]); then account for damping reduces to a suitable selection of possible iteration procedures [7].

In difference from the above used iteration procedure with calculations of $f_1^{(s)}$ where $s = 0, 1, 2, ...$ is an order of collision iteration at successive substitution the perturbed function of the lower order $f_1^{(s-1)}$ into collision integral, that is in a sequence

$$\hat{Q}\left(f_1^{(s+1)}\right) \leftarrow \hat{Q}\left(f_1^{(s)}\right) \tag{43}$$

One can also use iteration process in the more comfortable manner (see [7])

$$\hat{Q}\left(f_1^{(s+1)}\right) \leftarrow \hat{Q}\left(f_1^{(s)}\right) \frac{f_1^{(s+1)}}{f_1^{(s)}} \tag{44}$$

where at assumed convergence of iterations, $f_1^{(s+1)}/f_1^{(s)} \to 1$ at $s \to \infty$. It is shown [7] that results of at least the first iteration at both processes, (43) and (44), practically coincide.

Real value damping solution can be represented in the form of expansions in damping overtones

$$E = \sum_{n>0} e^{-\delta x} \left[ E_n^{(s)} e^{in(\omega t - k_0 x)} \right] + C.C., \tag{45}$$

$$F = \sum_{n>0} e^{-\delta x} \left[ F_n^{(s)} e^{in(\omega t - k_0 x)} \right] + C.C., \tag{46}$$

where C.C. means "complex conjugated" value; $s$ is iteration order for collision term; $k_0$ is real value part of wave number; $\delta$ is damping decrement, where $k_0$, $\delta$ are defined correspondingly by relations (24), (34) or (37), (38), (40), (42).

Expedience of such expansions is verified with the possibility of recurrent relations for $E_n^{(s)}$, $F_n^{(s)}$ at equalizing summed amplitudes of overtones with the same power exponents correspondingly either $\exp\left[-\delta x + in(\omega t - k_0 x)\right]$ or $\exp\left[-\delta x - in(\omega t - k_0 x)\right]$ in the left and right parts of wave equations (1), (2) and (3), (4).

For the first iteration order $n = 1$ one obtains already presented before equations for longitudinal and transverse waves correspondingly with the before presented expressions for $k_0$ and coordinate damping decrements $\delta$ in the first order iteration $s = 0$. The substitution of expansions (45), (46) into (1) … (4) leads to the expressions for overtones of the first and second orders $n = 1$ and $n = 2$:

*Longitudinal waves*

$$F_1^{(1)} \approx \frac{-ie}{m} E_1 \frac{\partial f_0/\partial v_x}{\omega - k_0 v_x + i\delta v_x + i\frac{\hat{Q}\left(F_1^{(0)}\right)}{F_1^{(0)}}}, \tag{47}$$

$$F_2^{(1)} \approx -\frac{ie}{m} \frac{\frac{\partial}{\partial v_x}\left(E_2^{(1)} f_0 + E_1 F_1^{(1)}\right)}{2\omega - 2k_0 v_x + i\delta v_x + i\frac{\hat{Q}\left(F_2^{(0)}\right)}{F_2^{(0)}}}, \tag{48}$$

$$E_2^{(1)} \approx \frac{-\omega_L^2}{2k_0 - i\delta} \int \frac{\frac{\partial}{\partial v_x}\left(E_2^{(1)} f_0 + E_1 F_1^{(1)}\right)}{2\omega - 2k_0 v_x + i\delta v_x + i\frac{\hat{Q}\left(F_2^{(0)}\right)}{F_2^{(0)}}} d\vec{v} . \tag{49}$$

*Transverse waves*

$$F_1^{(1)} \approx -\frac{ie}{m} E_1 \frac{\partial f_0/\partial v_z}{\omega - k_0 v_x + i\delta v_x + i\frac{\hat{Q}(F_1^{(0)})}{F_1^{(0)}}}, \qquad (50)$$

$$F_2^{(1)} \approx -\frac{ie}{m} \frac{\frac{\partial}{\partial v_z}\left(E_2^1 f_0 + E_1 F_1^{(0)}\right)}{2\omega - 2k_0 v_x + i\delta v_x + i\frac{\hat{Q}(F_2^{(0)})}{F_2^{(0)}}}, \qquad (51)$$

$$E_2^{(1)} \approx \frac{-2\omega\omega_L^2/c^2}{\delta^2 + 4ik_0\delta - 4k_0^2 + 4\omega^2/c^2} \int \frac{\frac{\partial}{\partial v_z}\left(E_2^{(1)} f_0 + E_1 F_1^{(1)}\right)}{2\omega - 2k_0 v_x + i\delta v_x + i\frac{\hat{Q}(F_2^{(0)})}{F_2^{(0)}}} d\vec{v}. \qquad (52)$$

From Eqs. (45), (46) it follows that the presence of quadratic term in kinetic equation leads to appearance of a sum of overtones with n-order exponents, amplitude of every one contains decreasing terms proportional to $E_1^n$, $E_1^{n+2}$, $E_1^{n+4}$, ... correspondingly to positive and negative values $q$, $r \neq 0$, $q + r = n$, $q + r \neq 0$.

Therefore for the précised calculation of overtones amplitude with recurrent formulas one can use iteration method précising successively amplitudes $\sim E_n$, $\sim F_n$ of the lower orders $n$ with the repeated calculations by the same formulas after successive addition there the terms with higher powers $(E_1)^{|q|+|r|}$, $|q|+|r| > 0$, but with the same order $q + r = n$, what results from cross-product of the two series of complex conjugated overtones.

Accounting for these notes one can write recurrent formulas of the higher orders $E_n$, $F_n$, however resulting expressions are very cumbersome and hardly can be useful already at $n > 3$.

Both iteration processes: calculation of collision term by iterations in $s$ as well as iterations in $n$ at calculations of overtone amplitudes with addition terms $\sim (E_1)^{|q|+|r|}$, $|q|+|r| > n$, appear independent. Considerable difficulties at numerical calculations can be tied to the considered in [8] overtones cut off near peculiar points of violation of kinetic equation where the full distribution function $f_0 + f_1$ becomes negative, and with the following renormalization of cut distribution functions.

**Conclusion**

We have presented a general form of solutions of equations of longitudinal as well as transverse electron waves in low-collision Maxwellian plasmas in the form of expansion in overtones of exciting boundary field frequency. There are presented recurrent relations for overtone amplitudes, there are considered iteration procedures for successive obtaining more precise values at following iterative inclusion of before neglected small terms $\sim (E_1)^{|q|+|r|}$, $q + r = n.$

The absence of non-physical exponentially divergent waves as well as of "Landau damping" is owing to refuse from the proposed by Landau way of calculation logarithmically divergent integrals with integration them along the contour, removed into complex values plane of integrand variant real value $v_x$, but instead to integrate along the real value axis $v_x$ in the principal value sense.

Solutions of wave equations for plasma with slight excess over Maxwellian distribution in the distribution tail are exponentially damping [4], [10]. The waves in collisionless Maxwellian plasma with produced by them weak overtones are non-damping.

In the case of low-collision plasma there arises very strong damping with the decrement $\delta \to \infty$ at $\omega \to \omega_L + 0$ (but excluding low-velocity electron waves with its peculiar damping).

The obtained relations open way to the more precise and complete calculations which however appear very complicated owing to difficulties of calculations the principle value sense integrals as well as successively carried out iteration procedures in collision term $\hat{Q}\left(F_n^{(s)}\right)$ and overtone amplitudes $\sim E_n^{(s)}$, $\sim F_n^{(s)}$ correspondingly to further accounting for the terms of higher order $\sim (E_1)^{|q|+|r|}$ with $q+r=n$. An additional difficulty is tied to necessity of cutting off overtones in peculiar points with negative distribution function $f_0 + f_1$, where the kinetic equation is violated, and renormalization procedures.

Results of the carried out analysis appear enough characteristic and unexpected for to be tested experimentally (e.g. presence of weak overtones and others).

Experimental producing of collisionless or low-collision plasma with Maxwellian distribution function appears extremely difficult problem accounting for the wall constraints and some assumed electron streams as effect of the discharge supporting external constant electric field.

In the case of a stream of Maxwellian plasma, characteristic velocity $v_{xeff}$ decreases, what corresponds to decreasing value $\operatorname{Re} k$ and increasing wave length. According to general considerations in [4] one might assume that at increasing stream velocity the possibility arises of appearance damping wave solutions, but this assumption requires some special detailed investigation.

Possibly, one of the most suitable object for applying this theory might be space plasma in the form of rarefied clouds of ionized hydrogen.

In conclusion, one ought to note once again, in order to avoid some misunderstanding that the original wave equations (1) … , (4) have infinite number of mathematically correct asymptotical solutions in dependence on the taken sense of indefinite integrals in these equations satisfying original equations with different characterized by these solutions physical features. One might note in this connection also an interesting possibility of the so-called peculiar solutions of differential equations of the type described in [11], even at the given initial and boundary conditions. The wave equations are identically satisfied at substitution of these solutions. However selection of the only solution, besides correspondence to the boundary and initial conditions, is defined by some additional purely physical requirements, e.g., correspondingly to a concrete problem: finiteness, stability, real value solution at real value initial and boundary conditions, correspondence to the physical sense and definition of physical values in original equations, and other possible physical features and requirements.